\input epsf
\epsfverbosetrue
\documentstyle[twocolumn,aps,prl]{revtex}
\begin{document}
\title{Re-entrance of the Metallic Conductance in a Mesoscopic Proximity 
Superconductor}
\author{P. Charlat$^1$, H. Courtois$^1$, Ph. Gandit$^1$, D. Mailly$^2$, 
A.F. Volkov$^3$, and B. Pannetier$^1$}
\address{$^1$ Centre de Recherches sur les Tr\`es Basses 
Temp\'eratures-C.N.R.S. associ\'e \`a l'Universit\'e Joseph Fourier, 25 Av. 
des Martyrs, 38042 Grenoble, France} 
\address{$^2$ Laboratoire de Microstructures et de 
Micro\'electronique-C.N.R.S., 196 Av. H. Ravera, 92220 Bagneux, 
France}
\address{$^3$ Institute of Radio Engineering and Electronics, Russian 
Academy of Sciences, Mokhovaya St. 11, 103907 Moscow, Russia}
\maketitle
\begin{abstract}
We present an experimental study of the diffusive transport in a normal metal 
near a superconducting interface, showing the re-entrance of the metallic 
conductance at very low temperature. This new mesoscopic regime comes in 
when the thermal coherence length of the electron pairs exceeds the sample 
size. This re-entrance is suppressed by a bias voltage given by the Thouless 
energy and can be strongly enhanced by an Aharonov Bohm flux. 
Experimental results are well described by the linearized quasiclassical 
theory.\end{abstract} 
\bigskip

\vspace{- 0.7 cm}
During the last few years, the proximity effect between a superconductor (S) 
and a 
normal (N) metal has met a noticeable revival, thanks to spectacular progress 
in the 
fabrication of samples of mesoscopic size \cite{NS}. Experimental study of 
the transport 
near a S-N interface has shown that the proximity effect strongly affects 
electron transport 
in mesoscopic S-N systems : the deviation $\Delta G$ of the conductance 
from 
its normal-state value strongly depends on temperature $T$ and oscillates in 
an applied
magnetic field $H$ if a N loop is present 
\cite{Petrashov,Dimoulas,CourtoisPrl}. 
Various theoretical approaches were suggested to explain this behavior. A 
scattering matrix method based upon Landauer formula \cite{Beenakker} as 
well as a 
numerical solution of the Bogolubov-de Gennes equations \cite{Lambert} 
were used. 
These studies demonstrated that superconductivity does not affect the charge 
transfer in 
the N metal if the temperature $T$ and the voltage $V$ are zero, i.e. $\Delta 
G$ is zero at 
zero energy. A more powerful method based on the equations for the 
quasiclassical 
Green's functions \cite{Volkov,Zaikin,Nazarov96,Volkov-Lambert} was 
used to obtain 
the dependence of $\Delta G$ on $T$ and $V$. It has been established 
\cite{Nazarov96} 
that at $V = 0$ the deviation of the conductance $\Delta G$ increases from 
zero at $T = 
0$ (if electron-electron interaction in N is negligeable) with increasing $T$, 
reaches a 
maximum at approximately the Thouless temperature $\epsilon_c/k_B = \hbar 
D/k_B 
L^2$ and decreases to zero at $T >> \epsilon_c/k_B$. This constitutes the 
re-entrance 
effect for the metallic conductance of the N metal. Similar dependence of 
$\Delta G(V)$ at 
$T = 0$ has been found in \cite{Volkov-Lambert} both in a numerical 
solution of the 
Bogoliubov-de Gennes equations and in a analytical solution of the equations 
for the 
quasiclassical Green's functions.

The physics behind this re-entrance effect involves non equilibrium effects 
between 
quasiparticles injected by the N reservoirs and electron pairs leaking from S. 
At the N-S 
interface, an incident electron is reflected into a hole of the same energy 
$\epsilon$ 
compared to the Fermi level $E_F$, but with a slight change in wave-vector 
$\delta k$ 
due to the branch crossing : $\delta k / k_F = \epsilon/E_F$, $k_F$ being the 
Fermi 
wavevector. The phase conjugation between the electron and the hole results 
in 
a finite pair amplitude involving states $(k_F+ \epsilon / \hbar v_F,-k_F+ 
\epsilon / \hbar 
v_F)$, $v_F$ being the Fermi velocity. Such a pair maintains coherence in N 
up to the 
energy-dependent diffusion length $L_\epsilon = \sqrt{\hbar D/\epsilon}$ 
\cite{Volkov,Zhou} which coincides with the well-known thermal length 
$L_T =\sqrt{\hbar D/k_B T}$ at $\epsilon = 2 \pi k_B T$.

\begin{figure}
\epsfxsize=7 cm \epsfbox{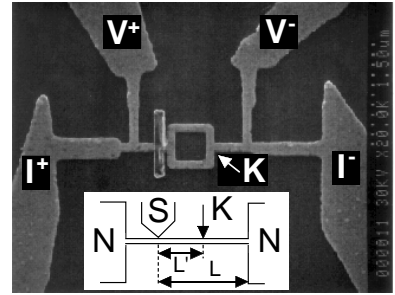}
\vspace{0.6 cm}
%\hspace{1 cm}
\caption{Micrograph of the sample made of a continuous Cu (= N) loop in 
contact with a 
single Al (= S) island. Inset : simple model. At half integer magnetic flux, a 
condition 
$F=0$ is enforced at the point K at $L'=$ 1 
$\mu m$ from S. The distances between S and the right and left N reservoirs 
are 
respectively $L=$ 2 $\mu m$ and 0.5 $\mu m$.}
\label{Photo}
\end{figure}

In the high-temperature regime $L_T < 
L$ or equivalently $\epsilon_c < k_B T$, it is well-known that proximity 
effect results in 
the subtraction of a length $L_T$ of N metal from the resistance of a S-N 
junction. In the 
low temperature $L_T > L$ or $\epsilon_c > k_B T$ and low voltage $eV < 
\epsilon_c$, electron pairs are coherent over the whole sample. The proximity 
effect on the N metal resistance is still predicted to be zero. In 
this Letter, we report the experimental realization of both limits ($L < L_T$ 
and $L > 
L_T$) and the observation of the re-entrance of the metallic conductance in a 
mesoscopic 
proximity superconductor. The low-temperature re-entrant regime is 
destroyed by 
increasing the temperature \cite{Nazarov96} or the voltage 
\cite{Volkov-Lambert}. As 
will be discussed below, an Aharonov-Bohm flux modifies the effective 
length of the 
sample and therefore shifts the energy crossover of the re-entrant regime.

Fig. \ref {Photo} shows a micrograph of the sample made of a square copper 
(Cu) loop 
in contact with a single aluminum (Al) island. The loop, although not 
essential for the 
occurrence of the re-entrance effect allows one to control boundary conditions 
for the pair 
amplitude. The Cu wire width is 150 nm and its thickness is about 40 nm. 
The distance 
between the Cu loop and the Al island is about 100 nm, whereas the perimeter 
of the loop is 2 $\mu m$. One should 
note that the sample geometry differs from all previous sample geometries 
with two 
superconducting contacts \cite{Petrashov,Dimoulas,CourtoisPrl} in that there 
is a single 
superconducting phase and therefore no possible Josephson contribution. 
Two voltage 
probes measure the distribution at the outflows of the reservoirs, which are 
the wide 
contact pads at both ends of the Cu wire. The Cu surface is in-situ cleaned 
before Al 
deposition in order to ensure an optimum transparency of the Cu/Al interface 
\cite{CourtoisPrl}.

\begin{figure}
\epsfxsize=8.5 cm \epsfbox{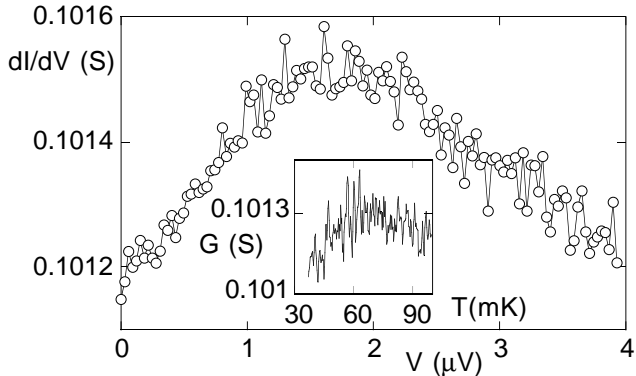}
\caption{Non monotoneous voltage dependence of the sample conductance at 
$T$ = 22 
mK ; the normal-state conductance $G_N$ is 0.0937 S. Inset: Temperature 
dependence. 
Measurement current of 70 nA in both curves.}
\label{Reentrance}
\end{figure}

We performed transport measurements in a $\mu$-metal--shielded dilution 
refrigerator 
down to 20 mK. From the normal-state conductance $G_{N}$ = 0.0937 S 
we find a 
diffusion coefficient $D =$ 70 $cm^2/s$, an elastic mean free path $l_e =$ 13 
nm and a 
thermal coherence length $L_T =$ 92 nm$/ \sqrt{T}$. Aluminum islands 
become 
superconducting below $T_c \simeq$ 1.4 K. The behaviour of the 
conductance in the 
high-temperature regime $L_T < L$ (i.e. above 500 mK) is very similar to 
the two-island 
case \cite{CourtoisPrl}. At 
lower temperatures, so that $L_T \simeq L$, and zero magnetic field, we 
observe a {\it 
decrease} of the low-voltage conductance (Fig. 2 inset). This occurs below 
50 mK, at the 
temperature where a Josephson coupling would be expected in a two-islands 
geometry. 
The voltage dependence of the measured conductance shows the most striking 
behaviour 
i.e. an {\it increase} of the conductance when the bias voltage is {\it 
increased} (Fig. 2). 
This non-linear behaviour discards an interpretation in terms of weak 
localization, which 
is known to be insensitive to voltage. The conductance peak is observed at a 
bias voltage 
(about 1.7 $\mu V$) of the order of the calculated Thouless energy 
$\epsilon_c =$ 1.1 
$\mu V$ related with a sample length $L =$ 2 $\mu m$. In Fig. 2 inset, the 
peak 
position is also consistent with the Thouless temperature $\epsilon_{c} / k_B 
=$ 13 
mK. One can note that the discussed energies are much smaller than Al 
energy gap 
$\Delta$.

\begin{figure}
\epsfxsize=8.5 cm \epsfbox{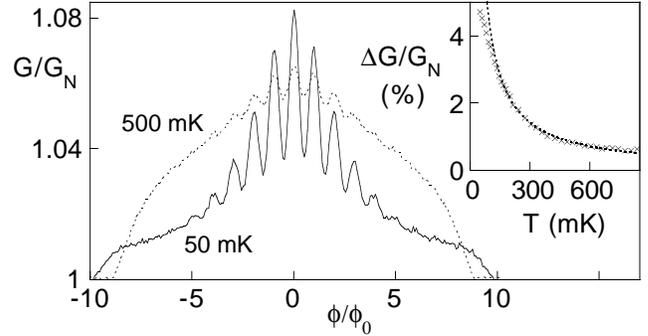}
\caption{Magnetoconductance oscillations showing $\Phi_0 = h/2e$ 
flux-periodicity at 
$T=$ 50 mK (solid line) and at $T=$ 500 mK (dashed). Re-entrance of the 
resistance is 
visible at half integer magnetic flux and at high field. Inset : oscillations 
amplitude together with a 1/T power law fit.}
\label{Osc.}
\end{figure}

Let us now analyze the effect of the magnetic field. Fig. 3 shows oscillations 
of the 
magnetoconductance with a periodicity of 
one flux quantum $\phi_0 = h/2e$ in the loop area. Here the re-entrance effect 
can be very 
clearly seen at $\phi = \phi_0 /2, 3\phi_0/2$ and at higher field. As previously 
observed 
in two-island samples, the oscillation amplitude decays slowly with a 
$1/T$--power-law 
down to 200 mK (Fig. 3 inset). Fig. 4 shows the temperature dependence of 
the 
conductance for various values of the magnetic flux in the loop. On this scale, 
the 
re-entrance at zero-field is hardly distinguishable \cite{LT}. At $\phi = \phi_0 
/ 2$, the 
conductance maximum is near 500 mK and the re-entrance has a much 
larger amplitude. At $\phi = \phi_0$, the curve is close to the zero-field case, 
and at $\phi 
= 3 \phi_0 /2$, close to the $\phi_0 / 2$ case. Increasing further the magnetic 
field (Fig. 4 
left part), the conductance peak is displaced to higher temperature and the 
$\phi_0$-periodic modulation is suppressed. Fig. 5 shows the conductance 
peaks 
energies obtained from both the current-voltage characteristics and 
temperature 
dependence of the conductance as a function of the magnetic flux. Hence the 
magnetic 
field has two effects on the re-entrance : (i) a large $\phi_0$ oscillation of the 
peak 
position at low field ; (ii) a monotonous shift at higher fields.

Most of our observations can be analysed in the framework of the 
quasiclassical theory 
for inhomogenous superconductors 
\cite{Volkov,Nazarov96,Volkov-Lambert,Zhou}. We present here a 
simplified 
version that however keeps the essential physical features. We consider the 
mesoscopic 
regime where the inelastic scattering length is larger than the sample length 
$L$ between 
the reservoirs. The flow of electrons at a particular energy $\epsilon$ is then 
uniform over 
the sample, and one has to consider transport through independent channels at 
the energy $\epsilon$. In a perfect 
reservoir, electrons follow the Fermi equilibrium distribution at the 
temperature $T$ and 
chemical potential $\mu$. Charges are injected in the system from one 
reservoir at $\mu = 
eV$ and transferred to the other, so that the current is carried by electrons 
within an 
energy window $[0,\mu = eV]$ with a thermal broadening $k_B T$. Let us 
assume that 
proximity effect can be accounted for as a conductivity enhancement $\delta 
\sigma(\epsilon,x)$ depending on both the energy $\epsilon$ and the distance 
$x$ from 
the S interface. From the behaviour of $\delta \sigma(\epsilon,x)$ it is then 
straightforward to calculate the excess conductance $\delta g(\epsilon )$ for 
the precise 
geometry of the sample. The excess conductance $\delta G(V,T)$ at voltage V 
and temperature T writes :
\begin {equation}
\delta G(V,T) = \int_{-\infty}^{\infty}\delta g(\epsilon ) P(V-\epsilon)d 
\epsilon 
\label {total}
\end {equation}
where $P(\epsilon)=[4k_B T ${cosh}$^2(\epsilon/2k_B T)]^{-1}$ is a 
thermal kernel 
which reduces to the Dirac function at $T = 0$. Hence, the low-temperature 
differential 
conductance $dI/dV = G_{N}+\delta G$ probes the proximity-induced 
excess 
conductance $\delta g(\epsilon)$ at energy $\epsilon = eV$ with a thermal 
broadening 
$k_{B}T$. Independent measurements of the excess conductance as a 
function of $T$ 
and $V$ agree with Eq. 1 for not too low temperatures and voltages, 
provided the 
chemical potential in the reservoirs is taken as equal to the measured voltage 
times a 
geometrical factor of 1.05. This correspondence however fails for energies 
less than 
about 5 $\mu eV$ : we believe that at these energies the inelastic 
collision rate is too small to ensure a good thermalisation in the reservoirs 
\cite{Charlat}.

In N and at zero magnetic field, $F(\epsilon,x)$ follows the Usadel equation 
that for 
small $F$ can be linearized as : 
\begin{equation} 
\hbar D \partial^{2}_{x} F + (2 i\epsilon - \frac{\hbar D}{L_{\varphi}^2}) F 
= 0
\label{diffusion}
\end{equation}
The linearization used above is a very crude approximation, since near the 
interface which 
is believed to be clean, the pair amplitude should be large (in this 
approximation, we have 
$F(\epsilon \ll \Delta,0) =-i\pi /2$ for a perfectly transparent interface 
\cite{Charlat}). At 
the contact with a N reservoir $F$ is assumed to be zero. However, this 
simple 
formulation enables a straightforward understanding of the physical root of 
the 
conductance enhancement in a proximity system. Indeed Eq. 2 features 
simply a diffusion 
equation for the pair amplitude $F$ at the energy $\epsilon$ with a decay 
length 
$L_\epsilon$ and a cut-off at $L_\varphi$. At a particular energy $\epsilon$, 
the real part 
of the pair amplitude is zero at the S interface, maximum at a distance 
$L_\epsilon$ if 
$L_\epsilon \ll L_\varphi,L $ and then decays in an oscillating way. The pair 
amplitude 
$F$ is responsible for the local enhancement of the conductivity $\delta 
\sigma(\epsilon,x) 
= \sigma_N (Re [F(\epsilon,x)])^{2}$ for small $F$, $\sigma_N$ being the 
normal-state 
conductivity \cite{Volkov,Nazarov96,Volkov-Lambert,Zhou}. It involves 
two 
contributions : a positive and dominant one which is similar to the 
Maki-Thompson 
fluctuation term in superconductors above $T_c$ \cite{MT} and a negative 
one related to 
the decrease of the density of states. The two contributions cancel each other 
at zero 
energy \cite{Volkov-Arcs}.

\begin{figure}
\epsfxsize=8.5 cm \epsfbox{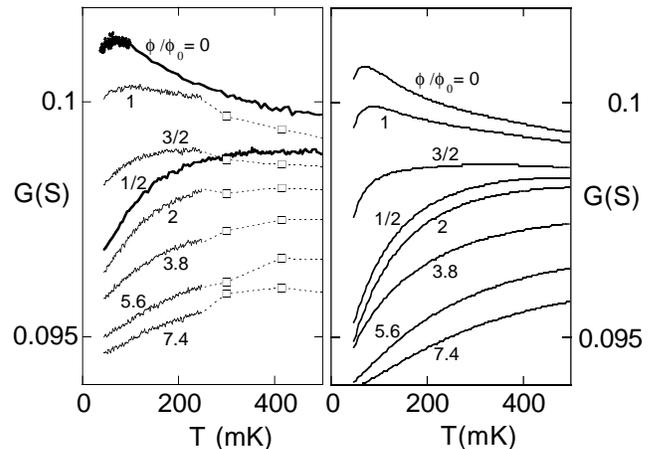}
\caption{Left : Measured temperature dependence of the conductance at 
different values of 
the magnetic flux $\Phi$ in units of the flux-quantum : $\Phi /\Phi _0$ = 0; 
1/2; 1; 3/2; 2; 
3.8; 5.6; 7.4. Measurement current is 200 nA. Right: Calculated conductance 
in the 
linear approximation for the sample model of Fig. 1 inset and at the same 
values of the 
flux. The only adjustable parameter is the effective width of the wires.}
\label{Champ-Calcul}
\end{figure}

We modelize the sample as two independent S-N circuits in series as shown 
in the inset 
of Fig. 1. This is our main approximation. It describes the main physics of 
our particular 
geometry and illustrates more general situations. Both circuits consist of a 
N-wire 
between a superconductor S and a normal reservoir N. Along the wire, the 
excess 
conductivity $\delta \sigma(\epsilon,x)$ at a given energy $\epsilon $ 
increases from zero 
at the N-S interface to a maximum of about 0.3 $\sigma_N $ at a distance 
$L_\epsilon$ 
from the interface (if $L \ll L_{\varphi}$) and then decays exponentially with 
$x$. The 
integrated excess conductance $\delta g(\epsilon)$ of the whole sample rises 
from zero 
with an ${\epsilon}^2$ law at low energy, reaches a maximum of 0.15 
$G_N$ at about 5 
$\epsilon_c$ and goes back to zero at higher energy with a 
$1/\sqrt{\epsilon}$ law.
This behaviour is indeed confirmed in the experimental results in Fig.2. We 
observe a 
conductance peak as function of both temperature and voltage. The 
conductance is 
maximum for a temperature (50 mK) close to the calculated crossover 
temperature ($5 
\epsilon_c / k_B =$ 65 mK), see Fig. 2 inset. Only qualitative agreement 
between the 
observed (1.7 $\mu V$) and calculated energy ($5 \epsilon_c =$ 
5.5 $\mu V$) is obtained. This discrepancy is believed to be due to 
insufficient energy 
relaxation efficiency at low energy in the Cu reservoirs.

Because of the loop geometry, a magnetic field induces an Aharonov-Bohm 
flux, which 
changes the boundary conditions on the pair amplitude $F(\epsilon )$. At zero 
magnetic 
flux, $F(\epsilon )$ is zero only at the contact with the normal reservoirs. At 
half 
magnetic flux, destructive interference of the pair functions in the two 
branches enforces a 
zero in $F(\epsilon )$ at the node K (see Fig. 1 inset). Consequently, the pair 
amplitude 
is also zero between the loop and the N reservoir. Half a flux-quantum then 
reduces the effective sample size to the length $L'$ between the S interface 
and the point 
K. In the intermediate temperature regime $k_{B}T > \epsilon_{c}$ (or 
$L_{T} < L)$, 
this modulates the conductance with a relative amplitude of the order of 
$\epsilon_{c}/k_{B}T$ \cite{Charlat}, in qualitative agreement with the 
experiment.

\begin{figure}
\epsfxsize=8.5 cm \epsfbox{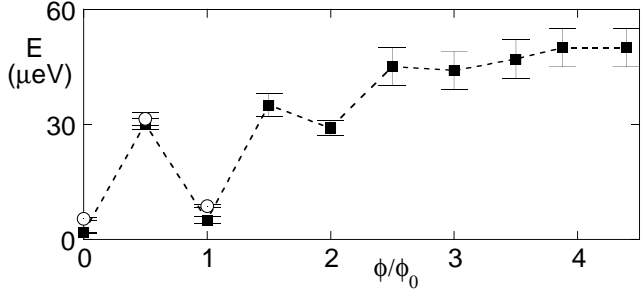}
\caption{Energy of the conductance maximum as a function of the magnetic 
flux $\phi$ in 
units of $\Phi_0$. The black square dots are obtained from the 
voltage-conductance 
$dI/dV(V)$ characteristics at $T=$ 100 mK, except for $\Phi /\Phi _0=0$ and 
$1$ 
($T=$22 and 45 mK). The white circles are obtained from Fig. 4. The 
discrepancies 
inbetween reflect the imperfections of the reservoirs at low energies.}
\label{Pics}
\end{figure}

As an additional effect of the magnetic field $H$, the phase-memory length 
$L_{\varphi}$ is renormalized due to the finite width $w$ of the Cu wire 
\cite{Pannetier-Rammal} :
\begin {equation}
L_{\varphi}^{-2}(H)=L_{\varphi}^{-2}(0)+\frac{\pi^{2}}{3} 
\frac{H^{2}w^{2}}{\Phi_{0}^{2}}
\end {equation}
When smaller than the sample length L, the phase-memory length 
$L_{\varphi}(H)$ 
plays the role of an effective length for the sample. As a result, the 
conductance peak is 
shifted to higher temperatures and energies when the magnetic field is 
increased, see Fig.4 and 5. At high magnetic field, the position of the 
conductance maxima does not 
increase as rapidly as could be expected because of the field-induced depletion 
of the gap 
$\Delta$. In the right part of Fig. 4, we show the calculated conductance 
using of Eq. 1-3 
in the modelized geometry of Fig.1 inset in the case of a fully transparent 
interface. The 
only free parameter is the width of the wires which has been adjusted so that 
the 
experimental damping of the amplitude of the magnetoconductance 
oscillations by the 
magnetic field, see Fig. 3, is well described by the calculation. The 
discrepancy between 
the fitted value $w=$ 65 nm and the measured value is attributed to deviations 
of sample 
geometry from our simple model. 
Our calculation accounts for both the global 
shape and amplitude of the curves and for their behaviour as a function of the 
magnetic 
flux. This is particularly remarkable in respect with the strong assumptions of 
the model. 
One should note that qualitative shape and amplitude of the curves are 
conserved if 
non-linearized Usadel equations or slightly different geometrical parameters 
are used.

In conclusion, we have measured the energy dependence of the proximity 
effect on the 
conductance near an N-S junction. As predicted in recent works 
\cite{Nazarov96,Volkov-Lambert}, we have observed the re-entrance of the 
metallic 
conductance 
when all energies involved are below the Thouless energy of the sample. In 
contrast with 
a very recent similar observation \cite{Hartog}, the energy crossover has 
been tracked as 
a function of temperature, voltage and magnetic field. Our 
experimental results are well described by the linearized Usadel equations 
from the 
quasiclassical theory.

We thank P. Butaud, M. D\'evoret, D. Est\`eve, B. Spivak, T. Stoof, A. 
Zaikin, and F. 
Zhou for stimulating discussions. A. F. V. thanks P. Monceau for 
hospitality, the 
Russian Fund for Fundamental Research and the 
collaboration program between the Ecole Normale Superieure de Paris and the 
Landau 
Institute for Theoritical Physics for support. We also acknowledge financial 
support from 
R\'egion Rh\^one-Alpes and D.R.E.T..
\vspace{- 0.5 cm}

\end{document}